\documentclass{elsart}
\usepackage{epsfig}

\newcommand{\be}{\begin{equation}}
\newcommand{\ee}{\end{equation}}
\newcommand{\beas}{\begin{eqnarray*}}
\newcommand{\eeas}{\end{eqnarray*}}
\newcommand{\bea}{\begin{eqnarray}}
\newcommand{\eea}{\end{eqnarray}}
\begin{document}
\begin{frontmatter}
\title{The Local Minority Game}
\author{S. Moelbert$^1$ and P. De Los Rios$^{1,2}$}
\address{$^1$Institut de Physique Th\'eorique,
Universit\'e de Lausanne, CH-1015, Lausanne, Switzerland.}
\address{$^2$INFM - Sezione di Torino, Politecnico di Torino, Corso Duca degli Abruzzi 24,
10129 Torino, Italy. }

\begin{abstract}
Ecologists and economists try to explain collective behavior in terms of
competitive systems of selfish individuals with the ability to learn from
the past. Statistical physicists have been investigating models which might
contribute to the understanding of the underlying mechanisms of these systems.
During the last three years one intuitive model, commonly referred to as the
Minority Game,
has attracted broad attention. Powerful yet simple, the minority game
has produced encouraging results which can explain the temporal behaviour
of competitive systems. Here we switch the interest to phenomena
due to a distribution of the individuals in space. For analyzing these effects we modify the
Minority Game and the Local Minority Game is introduced.
We study the system both numerically and analytically, using the customary techniques already developped
for the ordinary Minority Game.

\phantom{o}


\end{abstract}

\end{frontmatter}

In recent years statistical physicists have become increasingly interested
in phenomena of collective behaviour related to populations of interacting
individuals, typical of economic and biological systems. The aim is to find
models that are easy to handle, but still describe the self organization of
populations as accurately as possible. Such models can be found by applying the
tools of game theory, that analyze the mechanisms of individuals reasoning
in groups.
 Inspired by W. B. Arthur's Farol Bar Problem~\cite{arthur}, D. Challet and
Y.-C. Zhang proposed the minority game (MG)~\cite{challet1}: it is a simple,
but rich model describing a
population of selfish individuals fighting for a common resource, as
in the stock market, on roads or at waterholes. At a given time, each
individual chooses between two opposing actions, that can be simplified
as $a = \pm1$. In the above examples, these options might be to buy or to
sell, to take the highway or the backroad or to go to waterhole A or B.
Since overall resources are limited the aim of each individual is to make
the choice shared by the minority of the population. Each individual
does not know what the others will choose; consequently, there is in
general no best solution to the problem~\cite{game} and the individuals
take their decisions using strategies, that is rules for choosing
in each situation, based on the individual's own beliefs.
The only information on which an individual can base his decisions is
the history $\mu$ of the game, that is the sequence of the last $M$ winning
choices, e.g. $+1,+1,-1$ for $M=3$, where $M$ denotes the length of his
memory. A strategy $s$ then defines which action to take in every
situation, and hence consists in a sequence of $+1$s and $-1$s, one
for each possible history $\mu$. Since there are two possible choices,
the number of histories is $2^M$, which leads to a total number of
$2^{2^M}$ possible strategies, $\{S_i\}$.
In the original formulation of the MG, each individual
was assumed to possess two strategies. These strategies are chosen by each player
randomly from $\{S_i\}$, and never changed during the game.
In general the assigned strategies differ from individual to individual:
this reflects their different beliefs.
Individuals do not choose a specific action, but
rather a strategy that determines what action to take.
At every turn of the game the strategies that would have won get a reward, and
at the next turn the players choose the strategy with the highest score
(this rule can be strict, or probabilistic: the strategy with the highest score
has the greatest probability to be chosen).
By this mechanism (reinforcement learning), players are able to {\em learn} whether they
have a better strategy to use, and the system as a whole shows an extremely rich behavior.
%

The original MG has been studied extensively through numerical simulations,
with the introduction of many different features~\cite{challet2}, and then analytically solved
through a clever mapping onto a spin glass-like model~\cite{marsili}.
The exact solution shows that the MG undergoes a phase transition
between a symmetric phase, in which all the individuals perform as if they
were just acting randomly, and
an asymmetric phase where some individuals always play the same strategy
which turns out to be constantly the winning one. In this phase the system
is the most efficient, in the sense that on the average there is
always the possible maximum number of winners.
Moreover, it has been shown that in the stationary state
the histories may be replaced by random sequences without sensibly changing
the properties of the game~\cite{cavagna},
suggesting that all the histories occur with almost the same probability (the system is almost ergodic
in the history space).
So far, the system has been considered as a population of individuals,
each one interacting with all the others. In reality however an individual
is often in contact with only a small part of the whole population.
He has no information about how the members of his group interact
with members of other groups, and he tries to perform optimally in
his immediate surroundings.
 
In order to add this spatial aspect to the MG, we define a local
MG (LMG) (a different local version of the
MG has been introduced in ~\cite{Kalinowski00}).
Here, every individual still has the possibility to choose one of
his strategies
(for simplification, a number of $S=2$ strategies is assigned to
each individual) that dictates his next action. As in the global
game, each individual always tries to be in the minority group by
identifying the optimal strategy. The difference with respect to the MG
lies in
the definition of the region in which an individual is acting.
In the MG every individual was interacting with all the others; in the
LMG, instead, for each individual a small region is assigned, in
which he attempts to be in the minority.
To simplify, individuals occupy the sites of a $1d$ lattice, each one interacting
with his two nearest neighbours. A region thus contains $l=3$ individuals.
Each individual $i$ observes a different outcome $A_i$ of the game
than do the others and reconstructs a different history $\mu_i$.
His next action $a_i$ depends then
on his own subjective local history $\mu_i$.
The outcome $A_i(t)$ at time $t$ is defined as being the difference of the
number of individuals that chose $a=+1$ and the number of those that chose
$a=-1$. The sign of $A_i(t)$ then corresponds to the losing choice. Individual
$i$ knows the outcome in region $i$ at time $t$, given by
$A_i(t) = \sum_{j\in \Lambda_i} a_j^{\mu_j}(t)$ where $\Lambda_i$
denotes the set of individuals seen by player $i$. The reward for the players in the minority
is then proportional to $A_i$ (the fewer they are, the higher the reward).
The gain (or loss) individual $i$ makes at time $t$ can thus be defined as
\be
u_i(t) = - a_{i,s_i}^{\mu_i(t)}(t) A_i(t),
\label{gain}
\ee
where $s_i$
denotes the strategy he uses at time $t$, and his total profit up to $t$
is $U_i(t) = U_i(t-1) + u_i(t)$. The two strategies of individual $i$ are
indicated with $s_{i,1} = \uparrow$ and $s_{i,2} = \downarrow$. For both
his strategies individual $i$ knows the gain he would have made, if he
had used that strategy.
In accordance with its
score, a strategy $s_i$ is then chosen by individual $i$ with probability
\be \pi_{i,s_i}(t) = \frac{e^{\beta U_{i,s_i}(t)}}
{e^{\beta U_{i,\uparrow}(t)}+e^{\beta U_{i,\downarrow}(t)}},
\label{probs}
\ee
where $\beta$ plays the role of an inverse temperature $\beta = 1/T$.
For a system of $N$ individuals,
mean values can be written as
\be \langle A^{\mu} \rangle = \sum_{i=1}^N \sum_{s_i =\uparrow,
\downarrow} \pi_{i,s_i} a_{i,s_i}^{\mu_i}.
\ee
The convention of writing
$a_i^{\mu_i} = \omega_i^{\mu_i} + s_i \xi_i^{\mu_i}$,
as it was introduced in the MG~\cite{marsili}, leads to
\bea
\omega_i^{\mu_i} = \frac{a_{i,\uparrow}^{\mu_i} +
a_{i,\downarrow}^{\mu_i}}{2} & \:  \mbox{and } \: &
\xi_i^{\mu_i} = \frac{a_{i,\uparrow}^{\mu_i} -
a_{i,\downarrow}^{\mu_i}}{2}.
\eea
Defining $\Omega^{\mu_i} = \sum_{j \in \Lambda_i}{\omega_{j}^{\mu_j}}$
leads to the local outcome  $A^{\mu_i}
=\Omega^{\mu_{t}}+\sum_{j \in \Lambda_i}{\xi_{j}^{\mu_j}\,s_{j}}$.
The evolution of the mean value becomes
\begin{equation}
m_i(t) = \langle s_i \rangle =
\pi_{i,\uparrow} - \pi_{i,\downarrow} = \tanh \beta
\left[ U_{i,\uparrow}(t) - U_{i,\downarrow}(t)\right],
\label{moment}
\end{equation}
where for calculation we used $\uparrow = +1$ and
$\downarrow = -1$. In analogy to the MG~\cite{marsili}
the mean value is taken over all possible histories
$\mu_i$ for each individual. Taking the time derivative of (\ref{moment}),
and using (\ref{probs}),
after some simple algebra the dynamics become
\be
\frac{dm_i}{dt} = - 2 \beta \left[ 1-m_i^2 \right] \frac{dH }{dm_i},
\ee
where 
\be 
H= 2 \sum_{i=1}^{N} \overline{\Omega_i^{\mu_i} \xi_i^{\mu_i}} m_i +
\sum_{i=1}^{N} \sum_{j \in \Lambda_i}
\overline{\xi_i^{\mu_i} \xi_j^{\mu_j}} m_i m_j.
\label{hamiltonian}
\ee
$\overline{O}$ denotes the mean value of a quantity $O$ over all possible
histories of each region. A
spin-glass hamiltonian $H= \sum_i h_i m_i + \sum_i \sum_{j \in \Lambda_i}
J_{i,j} m_i m_j$ is thus found for spins
with values $m_i \in \left[ -1, +1 \right]$ in a
random field $h_i = \overline{\Omega_i^{\mu_i} \xi_i^{\mu_i}}$ and with
random
interactions $J_{i,j} = \overline{\xi_i^{\mu_i} \xi_j^{\mu_j}}$.
The system
minimizes $H$ for distributions $\{ J_{i,j} \}$, which are given by the
initial distribution of strategies among the individuals.
 
In the intuitive example of two anticorrelated strategies per individual, 
the action performed by strategy $\uparrow$ of individual $i$
is the opposite of the action performed by his strategy $\downarrow$ for
 every history, i.e.
$a_{i,\uparrow}^{\mu_i} = - a_{i,\downarrow}^{\mu_i}$. As a result,
$\omega_i^{\mu_i} = 0 \; \forall \; i, \mu_i$ and thus the field $h_i$
vanishes. In the simple case of $M=1$ the system behaves like an
antiferromagnet. This is straightforward, since the best
configurations are individuals taking alternative actions
($\uparrow \downarrow
\uparrow \downarrow \uparrow$ etc.). For $M=1$, chosing a strategy is
equivalent to choosing an action. For periodic boundary
conditions and an even number $N$ of individuals, everyone wins.
Analytically
this can be understood by calculating $J_{i,j}= \overline{\xi_i^{\mu_i}
\xi_j^{\mu_j}}$. On the one hand $\overline{\xi_i^{\mu_i} \xi_j^{\mu_j}} = 1
\; \forall \; i,j$ for which $a_{i, \uparrow}^{\mu_i} = a_{j,
\uparrow}^{\mu_j}$  and on the other hand  $\overline{\xi_i^{\mu_i}
\xi_j^{\mu_j}} = -1 \; \forall \; i,j$ for which $a_{i,
\uparrow}^{\mu_i} = a_{j, \downarrow}^{\mu_j}$, independent of the history.
This corresponds to an antiferromagnetic spin system at $T=0$.
The individuals build clusters in which everyone is
frozen to one of his strategies and therefore constantly wins. Between
two incompatible clusters there is a
boundary on which individuals are frustrated. Unlike the MG where the
maximum number of winners is $(N-1)/2$,
the LMG shows configurations where every individual can win, depending on
the nature of the regions. For
increasing $M$ the number of frozen individuals decreases until the system
reaches the state of random
individuals, a behavior which is confirmed by simulations. 
This example shows clearly how the LMG differs from the MG. In the MG,
the case
of two anticorrelated strategies is inefficient, which can be clearly understood
since it corresponds to an antiferromagnet in infinite dimensions, while in the LMG
it can lead to situations where everyone wins.
 
\begin{figure}
\centerline{\psfig{file=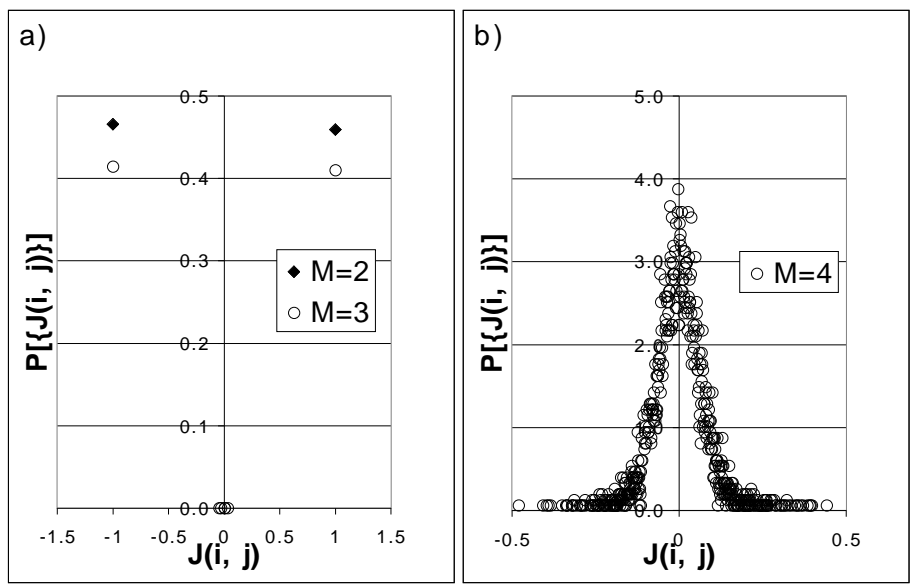,width=12.0cm}}
\caption{Distribution $P[\{J_{i,j}\}]$ at temperature $T=1.0$ for the
two dimensional LMG with nearest neighbor interaction and anticorrelated
strategies. a) For $M=2, 3$  the system is in the ordered phase, b)
while for $M=4$ it is in the disordered phase.}
\label{Fig1}
\end{figure}

The LMG is generalized to two dimensions by placing the individuals on a
square lattice with
periodic boundary conditions. Now each individual has four nearest
neighbours and the regions are chosen to
consist of $l=5$ individuals. Strategies are still taken as anticorrelated,
thus implying a vanishing random field $\{h_i\}$.
Analogous to the one-dimensional case, a mapping on a
spin system can be performed, which leads to the partition function
\be  Z[\{J_{i,j}\}] =
\int_{-1}^{+1} {\prod_i dm_i e^{-\beta H [ \{J_{i,j}, m_i, m_j\}]}}.
\label{partition}
\ee
In order to calculate the free energy, the average over the
distribution of the interactions $\{J_{i,j}\}$ needs to be taken. Still we do not know whether
we should take the average of the partition function itself (\ref{partition}) or of the free energy.
The choice depends on whether in (\ref{hamiltonian}) the interactions are annealed or quenched
random variables.
Numerical
simulations determine the distributions $P\left( J_{i,j} \right)$ to be
nearly gaussian at high temperatures, where the system is disordered ($m \simeq 0$; players
randomly switch between their strategies) and bimodal at low temperatures,
where almost all players are frozen ($|m| \simeq 1$; each player always plays the same stategy).
In the latter case $J_{i,j} = \pm 1$.
Since both the spins and the distribution of the interactions change during the dynamics,
the disorder is {\em annealed}, at variance with the ordinary MG, where disorder is quenched.
The origin of this behavior is in the dynamics in history space: whereas the
MG is (quasi)ergodic, so that the averages in (\ref{hamiltonian}) do not change in time,
the players of the LMG in the stationary state see just one single history,
and ergodicity is broken. Moreover they always choose a strategy that allows them
to be in the minority of their local environemnt, which leads to an effective aniferromagnet.
Temperature is not the only parameter affecting the distribution $P\left(J_{i,j}\right)$:
increasing $M$, a sudden
change from a bimodal to a gaussian distribution takes place (see Fig.\ref{Fig1}).

Actually, the dependence of the state of the system and of the distribution of the
interactions on $T$ and $M$ is not accidental.
The $\{m_i\}$ as well as the
interactions $\{J_{i,j}\}$ are free to take values which minimize the
Hamiltonian $H [ \{J_{i,j}, m_i, m_j\}]$ of the system, which therefore is
identified as an
annealed system with the partition function
\be Z_{\rm{ann}} =
\int_{-\infty}^{+\infty} \prod_{i, j \in \Lambda_i} dJ_{i,j} P[J_{i,j}]
\int_{-1}^{+1} {\prod_i dm_i e^{-\beta H [ \{J_{i,j}, m_i, m_j\}]}}.
\ee
Before thermalization ({\it i.e.} before the dynamics),
the distribution $P\left(J_{i,j}\right)$ is nearly gaussian (implying that all histories are equiprobable),
\be
P(J_{i,j}) = \frac{1}{\sqrt{2 \pi} \sigma_J}
e^{-\frac{J_{i,j}^2}{2 \sigma_J}}\;\;.
\ee
As a consequence, the hamiltonian of the
system can be written as
\be H_{\rm{ann}} [\{J_{i,j},m_i,m_j\}] = \sum_{i=1}^{N}
\sum_{j \in \Lambda_i}
\left( \frac{J_{i,j}^2}{2 \beta \sigma_J^2} -J_{i,j},m_i,m_j \right).
\ee
After an integration over the disorder the partition function becomes
\bea
Z_{\rm{ann}} & = & \int_{-1}^{+1} { \prod_{i} dm_i e^{ \frac{\beta^2
\sigma_J^2}{2} \sum_{i,j \in \Lambda_i} { m_i^2 m_j^2}}} \nonumber \\
&  =  &
\int_{-1}^{+1} { \prod_{i} dm_i e^{-\tilde{\beta} H_{eff}}}
\eea
where
$\tilde{\beta} \equiv \frac{ \beta^2 \sigma_J^2}{2}$ and
$H_{eff} = - \sum_{i,
j\in \Lambda_i} m_i^2 m_j^2$ is the effective hamiltonian, which no longer
depends explicitly on $\{J_{i,j}\}$. 

\begin{figure}
\centerline{\psfig{file=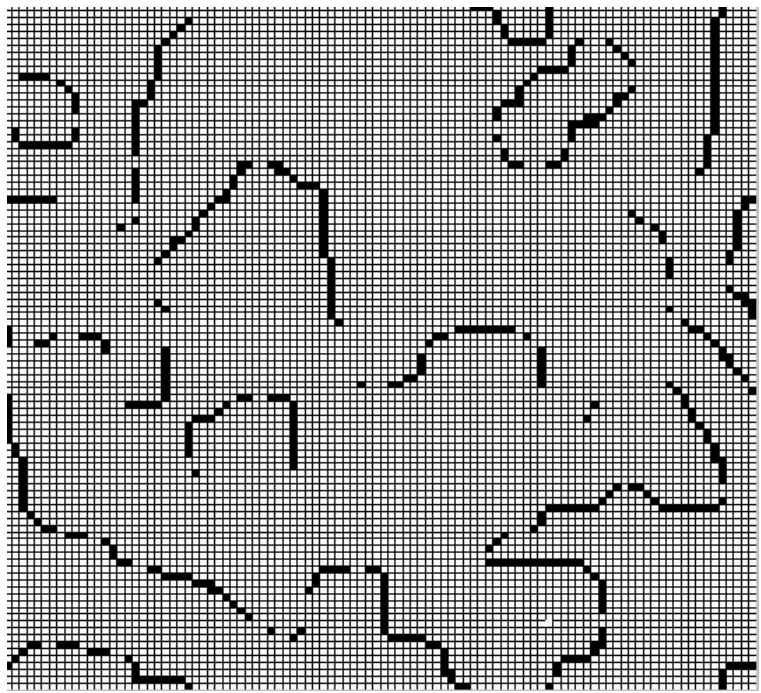,width=10.0cm}}
\caption{The system for $M=3$ and $T=1.$: white squares correspond to frozen players,
black squares to players that have not yet frozen.}
\label{Fig2}
\end{figure}

\begin{figure}
\centerline{\psfig{file=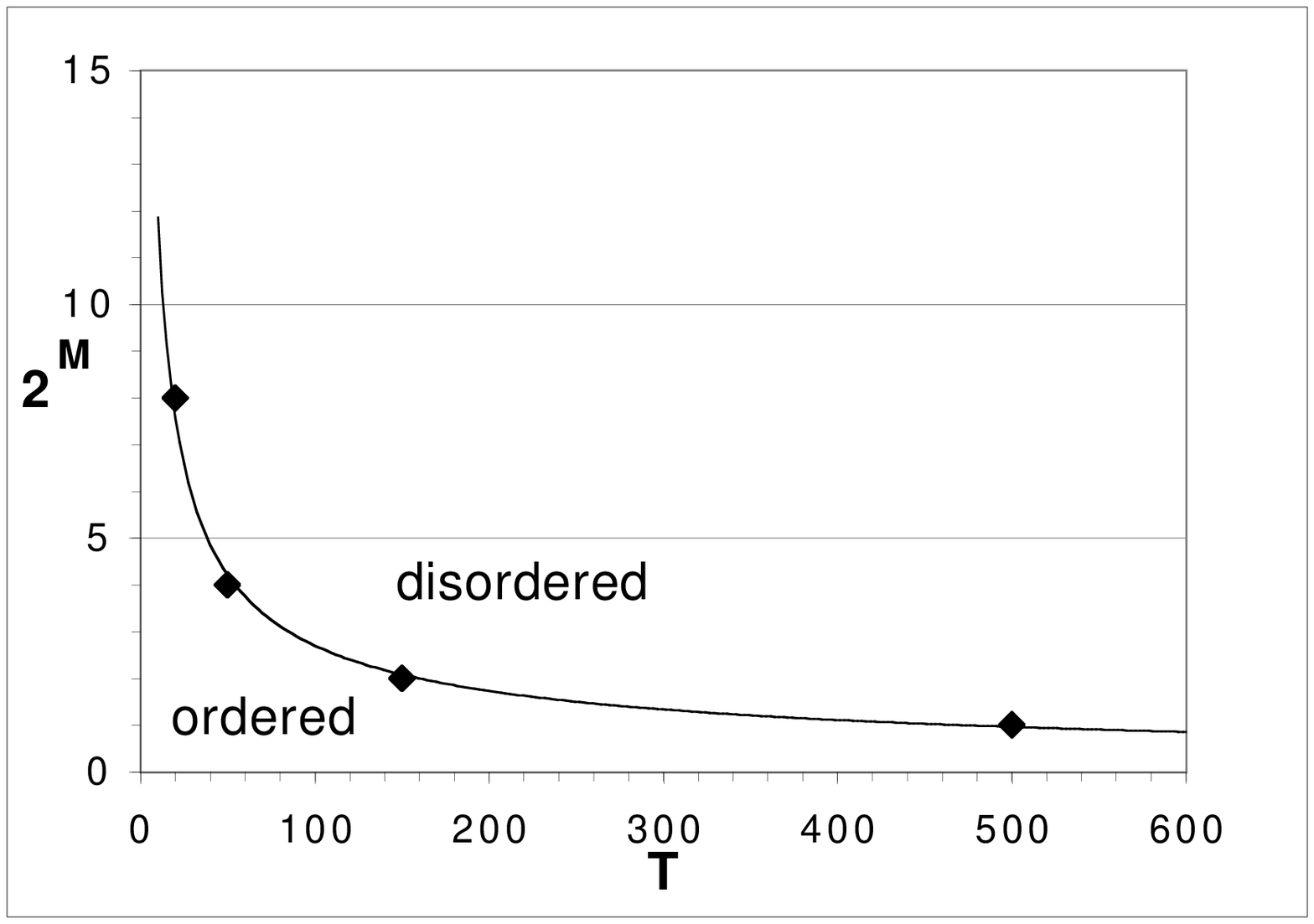,width=12.0cm}}
\caption{Phase diagram for the LMG with anticorrelated strategies in
two dimensions, where each individual interacts with his four nearest neighbors.
The system is in the ordered phase for low temperatures and low $M$.}
\label{FigX}
\end{figure}

At low temperatures the system tends to $<m_i^2> = 1$, while at high
temperatures $<m_i^2> = \int_{-1}^{+1}\frac{m_i^2}{2} dm_i = \frac{1}{3}$,
because in this case all possible values of $m_i$ occur with
uniform probability in $[-1,+1]$. In between, a critical temperature
$\tilde{T}_c$ is observed, which separates the two cases, and obeys $T_c^2
\propto \tilde{T}_c \sigma_J^2$. An approximative calculation
of the variance $\sigma_J^2 = <J_{I,j}^2>$ depending on $M$ confirms a
phase boundary at $T_c \sim \sigma_J \sim 2^{-M}$. This last result
can be obtained assuming that the histories of the neighbors $i$ and $j$
are independent (which of course is an approximation).
Fig.\ref{Fig1} shows the distribution $P[\{J_{i,j}\}]$ in the
stationary phase for three values of $M$ at a given temperature $T = 1.0$.
As can be seen, there is a sharp transition from a bimodal distribution
for $M=2$ and $M=3$ to a nearly gaussian distribution for $M=4$.
In Fig.\ref{Fig2} the stationary state of the system for $M=3$ and
$T=1.0$ is represented, clearly showing that the system is ordered (white
squares represent frozen players, black squares unfrozen ones),
with a bimodal distribution of the interactions.
For small values of $M$, all the individuals always use only one
of their strategies and win. The histories therefore do not change
over time, which clearly breaks the ergodicity of the dynamics in
history space. For $M=4$, however, the system is in the disorderd
phase and all the histories occur with roughly the same probability. This
confirms the calculated phase transition at a critical temperature
$T_c$ which is proportional to $\sigma_J \propto \frac{1}{2^M}$.
At higher temperature, the phase transition from the bimodal to
the gaussian distribution takes place at lower values of $M$.
The phase diagram is displayed in Fig.\ref{FigX}.

In the case of small regions, as considered in the above calculations, 
the influence of the individual agents on the outcome cannot be neglected;
indeed, if one
agent had chosen his unused strategy in the last step, the winning
choice could have easily been the other one. When an agent does not know
the effective impact of both his strategies in the last step, but only the
outcome while using one of them, he overestimates his second, unused
strategy
considerably~\cite{marsili2}. A clearly higher efficiency is thus observed
in a
population with complete information, meaning the agents know about the
influence that
both their strategies have on the outcome. In such a system nearly all
the agents win at low
temperature and only at very high temperature and for large $M$  does
decreasing efficiency occur. Even for $T=1000$ and $M=8$ the system
is clearly in the ordered phase with $99.9\%$  of all the individuals
frozen to one of their strategies.
If the histories are replaced by random sequences of winning choices,
the system with complete information is still more efficient than the one
with partial information, but not as clearly as for real histories.
The case of partial information, however, is the most realistic.

Although the basic principles of the LMG are the same as for the MG, the
behaviors of the two systems have considerable differences in
important features.
Unlike in the MG, where all the individuals interact with each other,
the interactions in the LMG are only local. The system
nevertheless shows a global, collective behavior, because it benefits
from the spatial arrangement of the individuals. The dynamics minimize a
hamiltonian that still depends on the $m_i$, as in the MG. Yet in the
LMG also the interactions change during the dynamics: the
disorder is annealed, while in the MG it is quenched.
Whereas the case of two anticorrelated strategies per individual is
inefficient in the MG, in the LMG it is highly efficient and there
are cases where all the individuals can win. On the other hand,
random strategies are disadvantageous in the LMG, because here every
individual has his own, subjective information. In a magnetic system language,
random strategies are in general incompatible with $J_{i,j} = \pm 1$, that is the
best interaction in the ordered case, and some residual disorder will still be
present even at low temperatures.
Moreover, random strategies give rise to a non-vanishing random
fields $h_i$. In two-dimensionsit is known that the random-field Ising model
is disordered down to $T=0$. Hence, we would need a three dimensional system
to recover the phase transition described above.

Our analysis has therefore shown, analytically and numerically,
that despite their similar structure, the MG and the LMG are significantly different:
whereas in the MG players organize in time, in the LMG space correlations become important.
It is concievable that reality sits in between these two extreme cases:
players could gather information both from their local environments and
from some distant sources: a small-world type of connections could be
a good modelization ofsuch a scenario; alternatively, histories could be formed
mixing both the local and the global information, with weigths mimicking the relevance of
each of them to the players. It is appealing to think that both time and space
coordinations could become important in such mixed games.

The authors thank the University of Fribourg (CH), where this work was
begun. This work has been supported by the Swiss National Science Foundation
under contract FNRS 21-61397.00 and by the EU Network
ERBFMRXCT980183.
A Java Applet to play with the LMG can be found at\newline
{\em http://www.unifr.ch/perso/moelbert/lmg\_nn\_fro/lmg\_nn\_hole\_not\_fro.html}


\begin{thebibliography}{99}

\bibitem{arthur} 
W.B. Arthur, Amer. Econ. Assoc. Papers and Proc. {\bf 84}, 406 (1994).

\bibitem{challet1} 
D. Challet and Y.-C. Zhang, Physica A {\bf 246}, 407 (1997). 

\bibitem{game}
D. Fudenberg and J. Tirole, {\em Game Theory}, (MIT Press, Cambridge, 1993);
J. W. Weibull, {\em Evolutionary Game Theory}, (MIT Press, Cambridge, 1995).

\bibitem{challet2}
D. Challet and Y.-C. Zhang, Physica A {\bf 256}, 514 (1998);
R. Savit, R. Manuca and R. Riolo, Phys. Rev. Lett. {\bf 82}, 2203 (1999).
N.F. Johnson, M. Hart and P.M. Hui, Physica A, {\bf 269}, 1 (1999);
R. D'hulst and G.J. Rodgers, Physica A, {\bf 270}, 222 (1999).

\bibitem{marsili}
D. Challet and M. Marsili, Phys. Rev. E {\bf 60}, R6271 (1999);
D. Challet, M. Marsili and R. Zecchina, Phys. Rev. Lett. {\bf 84}, 1824
(2000);
M. Marsili, D. Challet and R. Zecchina, Physica A {\bf280}, 522 (2000).


\bibitem{cavagna} 
A. Cavagna, Phys. Rev. E {\bf 59}, R3783 (1999).


\bibitem{Kalinowski00} T. Kalinowski, H.-J. Schultz and M. Briese, Physica A {\bf 277},
502 (2000).

\bibitem{marsili2}
D. Challet and M. Marsili, Phys. Rev. E {\bf 60}, R6271 (1999).

\bibitem{mpv} 
M. M\'ezard, G. Parisi and M. A. Virasoro, {\em Spin Glass Theory And Beyond}
(World Scientific, Singapore, 1986). 

\end{thebibliography}
\end{document}